\documentclass[a4paper,UKenglish,cleveref, autoref, thm-restate, authorcolumns]{lipics-v2021}
\usepackage[utf8]{inputenc}
\usepackage{CONCUR2021}

\nolinenumbers

\title{Formally Verified Simulations of State-Rich Processes using Interaction Trees in Isabelle/HOL}
\titlerunning{Formally Verified Simulations of State-Rich Processes using Interaction Trees}

\author{Simon Foster}{University of York}{simon.foster@york.ac.uk}{https://orcid.org/0000-0002-9889-9514}{EPSRC EP/S001190/1 (CyPhyAssure)}
\author{Chung-Kil Hur}{Seoul National University}{gil.hur@sf.snu.ac.kr}{}{}
\author{Jim Woodock}{University of York}{jim.woodcock@york.ac.uk}{https://orcid.org/0000-0001-7955-2702}{EPSRC EP/V026801/1 (TAS Verifiability), EP/M025756/1 (RoboCalc)}

\authorrunning{Simon Foster, Chung-Kil Hur, and Jim Woodcock}

\date{April 2021}



\begin{document}

\maketitle

\begin{abstract}
  Simulation and formal verification are important complementary techniques necessary in high assurance model-based systems development. In order to support coherent results, it is necessary to provide unifying semantics and automation for both activities. In this paper we apply Interaction Trees in Isabelle/HOL to produce a verification and simulation framework for state-rich process languages. We develop the core theory and verification techniques for Interaction Trees, use them to give a semantics to the CSP and \Circus languages, and formally link our new semantics with the failures-divergences semantic model. We also show how the Isabelle code generator can be used to generate verified executable simulations for reactive and concurrent programs.
\end{abstract}

\section{Introduction}

Simulation is an important technique for prototyping system models, which is widely used in several engineering domains, notably robotics and autonomous systems~\cite{Cavalcanti2019RoboSim}. For such high assurance systems, it is also necessary that controller software be formally verified, to ensure absence of faults. In order for results from simulation and formal verification to be used coherently, it is important that they are tied together using a unifying formal semantics.

Interaction trees (ITrees) have been introduced by Xia et al.~\cite{ITrees2019} as a semantic technique for reactive and concurrent programming, mechanised in the Coq theorem prover. They are coinductive structures, and therefore can model infinite behaviours supported by a variety of proof techniques. Moreover, ITrees are deterministic and executable structures and so they can provide a route to both verified simulators and implementations.

Previously, we have demonstrated an Isabelle-based theory library and verification tool for reactive systems~\cite{Foster17c,Foster2021-JLAMP}. This supports verification and step-wise development of nondeterministic and infinite state systems, based on the CSP~\cite{Brookes1984,Hoare85} and \Circus~\cite{Woodcock2001-Circus} process languages. This includes a specification mechanism, called reactive contracts, and calculational proof strategy. Extensions of our theory support reasoning about hybrid dynamical systems, which make it ideal for verifying autonomous robots. Recently, the set-based theory of CSP has also been mechanised~\cite{Taha2020CSP-Isabelle}. However, such reactive specifications, even if deterministic, are not executable and so there is a semantic gap with implementations.

In this paper, we demonstrate how ITrees can be used as a foundation for verification and simulation of state-rich concurrent systems. For this, we present a novel mechanisation of ITrees in Isabelle/HOL, which requires substantial adaptation from the original work. The benefit is access to Isabelle's powerful proof tools, notably the \skey{sledgehammer} automated theorem prover integration~\cite{Blanchette2016Hammers}, but also the variety of other tools we have created in Isabelle/UTP~\cite{Foster2020-IsabelleUTP}. Isabelle's code generator allows us to automatically produce ITree-based simulations, which allows a tight development loop, where simulation and verification activities are intertwined. All our results have been mechanised, and can be found in the accompanying repository\footnote{\url{https://github.com/isabelle-utp/interaction-trees}}, and specific icon links (\isalogo / \hasklogo) next to each result.

The structure of our paper is as follows. In \S\ref{sec:itrees} we show how ITrees are mechanised in Isabelle/HOL, including the core operators, and strong and weak bisimulation techniques. In \S\ref{sec:csp-circus} we show how deterministic CSP and \Circus processes can be semantically embedded into ITrees, including operators like external choice and parallel composition. In \S\ref{sec:sem-links} we link ITrees with the standard failures-divergences semantic model for CSP, which justifies their integration with other CSP-based techniques. In \S\ref{sec:simulation} we show how the code generator can be used to generate simulations. In \S\ref{sec:related} we briefly consider related work, and in \S\ref{sec:concl} we conclude.

\section{Interaction Trees in Isabelle/HOL}
\label{sec:itrees}

Here, we introduce Interaction Trees (ITrees) and develop the main theory in Isabelle/HOL, along with several novel results. ITrees were originally mechanised in Coq by Xia et al.~\cite{ITrees2019}. Our mechanisation in Isabelle/HOL brings unique advantages, including a flexible frontend syntax, an array of automated proof tools, and code generation to several languages.

ITrees are potentially infinite trees whose edges are decorated with events, representing the interactions between a process and its environment. They are parametrised over two sorts (types): $E$ of events and $R$ of return values (or states). There are three possible interactions: (1) termination, returning a value in $R$; (2) an internal event ($\tau$); or (3) a choice between several visible events. In Isabelle/HOL, we encode ITrees using a codatatype~\cite{Blanchette2014BNF,Blanchette2017Coinductive}: \isalink{https://github.com/isabelle-utp/interaction-trees/blob/ff9f73f98c653b265bd9da55689715cf973499c1/Interaction_Trees.thy\#L21}
\vspace{-1ex}
\begin{alltt}
  \isakwmaj{codatatype} ('e, 'r) itree = 
    Ret 'r | Sil "('e, 'r) itree" |  Vis "'e \(\pfun\) ('e, 'r) itree"
\end{alltt}
\vspace{-1ex}
Type parameters \texttt{'e} and \texttt{'r} encode the sorts $E$ and $R$. Constructor $\skey{Ret}$ represents a return value, and \skey{Sil} an internal event, which evolves to a further ITree. A visible event choice ($\skey{Vis}$) is represented by a partial function ($A \pfun B$) from events to ITrees, with a potentially infinite domain. This representation is the main deviation from ITrees in Coq~\cite{ITrees2019}. Here, $A \pfun B$ is isomorphic to $\texttt{A} \to \texttt{B option}$, where $\texttt{B option}$ can take the value \texttt{None} or \texttt{Some x} for \texttt{x::B}. We usually specify partial functions using $\lambda x \in A @ f(x)$, which restricts a function $f$ to the domain $A$. We write $\{\mapsto\}$ for an empty function, and adopt several operators from the Z notation~\cite{Spivey89}, such as $\dom$, override ($F \oplus G$), and domain restriction $(A \dres F)$.

We sometimes use $\ret{v}$ to denote $\Ret~v$, $\tau P$ to denote $\Sil~P$, and $\vbar\, e\!\in\!E \then P(e)$ to denote $\Vis (\lambda e \in E @ P(e))$. We write $e_1 \then P_1 \mathop{\vbar} \cdots \mathop{\vbar} e_n \then P_n$ when $E = \{e_1, \cdots, e_n\}$. We use $\tau^n P$ for an ITree prefixed by $n \in \nat$ internal events. We define $\skey{stop} \defs \Vis~\{\mapsto\}$, a deadlock situation where no event is possible. An example is $a \then \tau(\ret{x}) \mathop{\vbar} b \then \skey{stop}$, which can either perform an $a$ followed a $\tau$, and then terminate returning $x$, or perform a $b$ and then deadlocks.

We call an ITree \skey{unstable} if it has the form $\tau P$, and \skey{stable} otherwise. An ITree stabilises, written $\stabilises{P}$, if it becomes stable after a finite sequence of $\tau$ events, that is $\exists n~P' @ P = \tau^n P' \land \skey{stable}(P')$. An ITree that does not stabilise is divergent, written $\divergent{P} \defs \neg (P \Downarrow)$.

Using the operators mentioned so far, we can specify only ITrees of finite depth. Infinite interaction trees are specified using primitive corecursion~\cite{Blanchette2014BNF}, as exemplified below. \isalink{https://github.com/isabelle-utp/interaction-trees/blob/ff9f73f98c653b265bd9da55689715cf973499c1/ITree_Divergence.thy\#L77}
\vspace{-.5ex}
\begin{alltt}
  \isakwmaj{primcorec} div :: "('e, 's) itree" \isakwmin{where} "div = \(\tau\) div"
  \isakwmaj{primcorec} run :: "'e set \(\Rightarrow\) ('e, 's) itree" \isakwmin{where}
    "run E = Vis (map_pfun (\(\lambda\) x. run E) (pId_on E))"
\end{alltt}
\vspace{-.5ex}
The \isakwmaj{primcorec} command requires that every corecursive call on the right-hand side of an equation is guarded by a constructor. ITree \skey{div} represents the divergent ITree that does not terminate, and only performs internal activity. It is divergent, $\divergent{\skey{div}}$, since it never stabilises. Moreover, we can show that $\skey{div}$ is the unique fixed-point of $\tau^{n+1}$ for any $n \in \nat$, $\tau^{n+1} P = P \iff P = \skey{div}$, and consequently $\skey{div}$ is the only divergent ITree: $\divergent{P} \implies P = \skey{div}$.

ITree $\skey{run}~E$ can repeatedly perform any $e \in E$ without ceasing. It has the equivalent definition of $\skey{run}~E \defs \vbar e \in E \then \skey{run}~E$, and thus the special case $\skey{run}~\emptyset = \skey{stop}$. The formulation above uses the function $\texttt{map\_pfun} :: (\texttt{'b}\!\Rightarrow\!\texttt{'c}) \Rightarrow (\texttt{'a}\!\pfun\!\texttt{'b}) \Rightarrow (\texttt{'a}\!\pfun\!\texttt{'c})$ which maps a total function over every output of a partial function. Function \texttt{pId\_on E} is the identity partial function with domain \texttt{E}. This formulation is required to satisfy the syntactic guardedness requirements. For the sake of readability, we elide these details in the definitions that follow.

Corecursive definitions can have several equations ordered by priority, like a recursive function. We specify a monadic bind operator for ITrees using such a set of equations.

\begin{definition}[Interaction Tree Bind] \label{def:mbind} We fix $P, P' : (E, R)\skey{itree}$,
  $K : R \to (E, S)\skey{itree}$, $r : R$, and $F : E \pfun (E, S)\skey{itree}$. Then, $P \mbind K$
  is defined corecursively by the equations \isalink{https://github.com/isabelle-utp/interaction-trees/blob/ff9f73f98c653b265bd9da55689715cf973499c1/Interaction_Trees.thy\#L96}
  $$\ret{r} \mbind K = K~r \quad \tau P' \mbind K = \tau(P' \mbind K) \quad \Vis~F \mbind K = \Vis~(\lambda e \in \dom(F) @ F(x) \mbind K)$$
\end{definition}
The intuition of $P \mbind K$ is to execute $P$, and whenever it terminates ($\tick_x$), pass the given value $x$ on to the continuation $K$. We term $K$ a Kleisli tree, or KTree, since it is a Klesli lifting of an ITree. KTrees are of great importance for defining processes that depend on a previous state. For this, we define the type synonym $(E, S)\skey{htree} \defs (S \Rightarrow (E, S)\skey{itree})$ for a homogeneous KTree. We define the Kleisli composition operator $P \fatsemi Q \defs (\lambda x. P x \mbind Q)$, so symbolised because it is used as sequential composition. Bind satisfies several algebraic laws:

\begin{theorem}[Interaction Tree Bind Laws] \label{thm:bind-laws} $ $ \isalink{https://github.com/isabelle-utp/interaction-trees/blob/ff9f73f98c653b265bd9da55689715cf973499c1/Interaction_Trees.thy\#L178} 

  \vspace{-2ex}
  
    \begin{minipage}{.5\linewidth}
    \begin{align*}
        \Ret~x \mbind K &= K~x \\
        P \mbind \Ret &= P \\
        P \mbind (\lambda x. (Q~x \mbind R)) &= (P \mbind Q) \mbind R \\
        \skey{div} \mbind K &= \skey{div}
    \end{align*}
    \end{minipage}
    \begin{minipage}{.5\linewidth}
    \begin{align*}
        \Ret \fatsemi K &= K \\
        K \fatsemi \Ret &= K \\
        K_1 \fatsemi (K_2 \fatsemi K_3) &= (K_1 \fatsemi K_2) \fatsemi K_3 \\
        \skey{run}~E \mbind K &= \skey{run}~E
    \end{align*}
    \end{minipage}
\end{theorem}
Bind satisfies the three monad laws: it has $\Ret$ as left and right units, and is essentially associative. Moreover, both $\skey{div}$ and $\skey{run}$ are left annihilators for bind, since they do not terminate. From the monad laws, we can show that $(\fatsemi, \Ret)$ also forms a monoid. 

The laws of \cref{thm:bind-laws} are proved by coinduction, using the following derivation rule.

\begin{theorem}[ITree Coinduction] \label{thm:coind} We fix a relation $\mathcal{R} : (E, R)\skey{itree} \rel (E, R)\skey{itree}$ and then given $(P, Q) \in \mathcal{R}$ we can deduce $P = Q$ provided that the following conditions of $\mathcal{R}$ hold: \isalink{https://github.com/isabelle-utp/interaction-trees/blob/ff9f73f98c653b265bd9da55689715cf973499c1/Interaction_Trees.thy\#L63}
    $$\begin{array}{c}
        \forall (P', Q')\in \mathcal{R} @ \skey{is\_Ret}(P') = \skey{is\_Ret}(Q') \land \skey{is\_Sil}(P') = \skey{is\_Sil}(Q') \land \skey{is\_Vis}(P') = \skey{is\_Vis}(Q'); \\[.5ex]
        \forall (x, y) @ (\Ret~x, \Ret~y) \in \mathcal{R} \implies x = y; \forall (P', Q') @ (\Sil~P', \Sil~Q') \in \mathcal{R} \implies (P', Q') \in \mathcal{R}; \\[.5ex]
        \forall (F, G) @ (\Vis~F, \Vis~G) \in \mathcal{R} \implies (\dom(F) = \dom(G) \land (\forall e \in \dom(F) @ (F(x), G(x)) \in \mathcal{R}))
    \end{array}$$
\end{theorem}
To show $P = Q$, we need to construct a (strong) bisimulation $\mathcal{R}$ and show that $(P, Q) \in \mathcal{R}$. There are four provisos to show that $\mathcal{R}$ is a bisimulation. The first requires that only ITrees of the same kind are related, where $\skey{is\_Ret}$, $\skey{is\_Sil}$, and $\skey{is\_Vis}$ distinguish the three cases. The second proviso states that if $(\ret{x}, \ret{y}) \in \mathcal{R}$ then $x = y$. The third proviso states that internal events must yield bisimilar continuations: $(\tau P, \tau Q) \in \mathcal{R} \implies (P, Q) \in \mathcal{R}$. The final proviso states that for two visible interactions the two functions must have the same domain ($\dom(F) = \dom(G)$) and every event $e \in \dom(F)$ must lead to bisimilar continuations.

Next, we define an operator for iterating ITrees: \isalink{https://github.com/isabelle-utp/interaction-trees/blob/ff9f73f98c653b265bd9da55689715cf973499c1/ITree_Divergence.thy\#L352}
\begin{alltt}
\isakwmaj{corec} while :: "('s \(\Rightarrow\) bool) \(\Rightarrow\) ('e, 's) htree \(\Rightarrow\) ('e, 's) htree" \isakwmin{where}
"while b P s = (if (b s) then Sil (P s \(\mbind\) while b P) else Ret s)"
\end{alltt}
This is not primitively corecursive, since the corecursive call uses $\mbind$, and so we define it using the \isakwmaj{corec} command~\cite{Blanchette2015ExtCorec,Blanchette2017Corec} instead of \isakwmaj{primcorec}. This requires us to show that $\mbind$ is a ``friendly'' corecursive function~\cite{Blanchette2017Corec}: it consumes at most one input constructor to produce one output constructor. A while loop iterates whilst the condition $b$ is satisfied by state $s$. In this case, a $\tau$ event is followed by the loop body and the corecursive call. If the condition is false, the current state is returned. We introduce the special cases $\skey{loop}~F \defs \skey{while}~(\lambda s @ \skey{True})~F$ and $\skey{iter}~P \defs \skey{loop}~(\lambda s @ P)~()$, which represent infinite loops with and without state, respectively. We can show that $\skey{iter}~(\ret{()}) = \skey{div}$, since it never terminates and has no visible behaviour.

Though strong bisimulation is a useful equivalence, we often wish to abstract over $\tau$s. We therefore also introduce weak bisimulation, $P \approx Q$, as a coinductive-inductive predicate. It requires us to construct a relation $\mathcal{R}$ such that whenever $(P, Q)$ in $\mathcal{R}$ both stabilise, all their visible event continuations are also related by $\mathcal{R}$. For example, $\tau^m~P \approx \tau^n~Q$ whenever $P \approx Q$.  We have proved that $\approx$ is an equivalence relation, and $P \approx \skey{div} \implies P = \skey{div}$. \isalink{https://github.com/isabelle-utp/interaction-trees/blob/4bdea2d0a52341e7a19abc3950a3bcdd4b65e7fd/ITree_Weak_Bisim.thy}

\section{CSP and Circus}
\label{sec:csp-circus}

Here, we give an ITree semantics to deterministic fragments of the CSP~\cite{Brookes1984,Hoare85} and \Circus~\cite{Woodcock2001-Circus,Oliveira&09} languages. The standard CSP denotational semantics is provided by the failures-divergences model~\cite{Brookes1984,Roscoe2010-UCS}, and we provide preliminary results on linking to this in \S\ref{sec:sem-links}.

\subsection{CSP}

CSP processes are parametrised by an event alphabet ($\Sigma$), which specifies the possible ways a process communicates with its environment. For ITrees, $\Sigma$ is provided by the type parameter $E$. Whilst $E$ is typically infinite, it is usually expressed in terms of a finite set of channels, which can carry data of various types. Here, we characterise channels abstractly using prisms~\cite{Pickering2017-Optics}, a concept well known in the functional programming world:

\begin{definition}[Prisms] A prism is a quadruple $(\view, \Sigma, \pmatch, \pbuild)$ where $\view$
  and $\Sigma$ are non-empty sets. Functions $\pmatch : \Sigma \pfun \view$ and
  $\pbuild : \view \to \Sigma$ satisfy the following laws:
$$\pmatch(\pbuild~x) = x \qquad\quad y \in \dom(\pmatch) \implies \pbuild~(\pmatch~y) = y$$ We write $X : V \pto E$ if $X$ is a prism with $\Sigma_X = E$ and $\view_X = V$.
\end{definition}
Intuitively, a prism abstractly characterises a datatype constructor, $E$, taking a value of type $\mathcal{V}$. For CSP, each prism models a channel in $E$ carrying a value of type $\view$. We have created a command \isakwmaj{chantype}, which automates the creation of prism-based event alphabets.

CSP processes typically do not return data, though their components may, and so they are typically denoted as ITrees of type $(E, ())\skey{itree}$, returning the unit type $()$. An example is $\skey{skip} \defs \Ret~()$, which is a degenerate form of $\Ret$. We now define the basic CSP operators.

\begin{definition}[Basic CSP Constructs] \label{def:basic-constructs} \isalink{https://github.com/isabelle-utp/interaction-trees/blob/ff9f73f98c653b265bd9da55689715cf973499c1/ITree_CSP.thy\#L7}

  \vspace{-4ex}
  \begin{align*}
    \skey{inp} &:: (V \pto E) \Rightarrow V~\skey{set} \Rightarrow (E, V)\skey{itree} \\[-.5ex]
    \skey{inp}~c~A &\defs \Vis~(\lambda e \in \dom(\pmatch_c) \cap \pbuild_c \limg A \rimg @ \Ret~(\pmatch_c~e))
  \end{align*}

  \vspace{-4ex}
  
  \begin{minipage}{.5\linewidth}
  \begin{align*}
  \skey{outp} &:: (V \pto E) \Rightarrow V \Rightarrow (E, ())\skey{itree} \\[-.5ex]
    \skey{outp}~c~v &\defs \Vis~\{\skey{build}_c~v \mapsto \Ret~()\}
  \end{align*}
  \end{minipage}
  \begin{minipage}{.5\linewidth}
  \begin{align*}
    \skey{guard}~b &:: \mathbb{B} \Rightarrow (E, ())\skey{itree} \\[-.5ex]
    \skey{guard}~b &\defs (\textit{if}~b~\textit{then}~\skey{skip}~\textit{else}~\skey{stop})
  \end{align*}
\end{minipage}

\end{definition}
An input event ($\skey{inp}~c~A$) permits any event over the channel $c$, that is $e \in \dom(\pmatch_c)$, provided that its parameter is in $A$ ($e \in \pbuild_c \limg A \rimg$), and it returns the value received for use by a continuation. An output event ($\skey{outp}~c~v$) permits a single event, $v$ on channel $c$, and returns a null value of type $()$. We also define the special case $\skey{sync}~e \defs \skey{outp}~e~()$ for a basic event $e :: () \pto E$. A $\skey{guard}~b$ behaves as $\skey{skip}$ if $b = true$ and otherwise deadlocks. It corresponds to the guard in CSP, which can be defined as $b \guard P \defs (\skey{guard}~b \mbind (\lambda x @ P))$.

Using the monadic ``do'' notation, which boils down to applications of $\mbind$, we can now write simple reactive
programs such as $\skey{do} \{ x \leftarrow \skey{inp}~c; \skey{outp}~d~(2 \cdot x); \Ret~x \}$, which inputs $x$ over
channel $c : \nat \pto E$, outputs $2 \cdot x$ over channel $d$, and finally terminates, returning $x$.

Next, we define the external choice operator, $P \extchoice Q$, where the environment resolves the choice with an initial event of $P$ or $Q$. In CSP, $\extchoice$ can also introduce nondeterminism, for example $(a \then P) \extchoice (a \then Q)$ introduces an internal choice, since the $a$ event can lead to $P$ or $Q$, and is equal to $a \then (P \intchoice Q)$. Since we explicitly wish to avoid introducing such nondeterminism, we make a design choice to exclude this possibility by construction. There are other possibilities for handling nondeterminism in ITrees, which we consider in \S\ref{sec:concl}. As for $\mbind$, we define external choice corecursively using a set of ordered equations.

\begin{definition}[External choice] \label{def:extchoice} $P \extchoice Q$, is defined by the following set of equations: \isalink{https://github.com/isabelle-utp/interaction-trees/blob/ff9f73f98c653b265bd9da55689715cf973499c1/ITree_CSP.thy\#L75}

\vspace{-2ex}

\begin{minipage}{.35\linewidth}
\begin{align*}
    (\Vis~F) \extchoice (\Vis~G) &= \Vis~(F \odot G) \\
    (\Sil~P') \extchoice Q &= \Sil~(P' \extchoice Q) \\
    P \extchoice (\Sil~Q') &= \Sil~(P \extchoice Q')
\end{align*}
\end{minipage}
\begin{minipage}{.65\linewidth}
\begin{align*}
    (\Ret~x) \extchoice (\Vis~G) &= \Ret~x \\
    (\Vis~F) \extchoice (\Ret~y) &= \Ret~y \\
    (\Ret~x) \extchoice (\Ret~y) &= (\textit{if}~x = y~\textit{then}~(\Ret~x)~\textit{else}~\skey{stop})
\end{align*}
\end{minipage}

\vspace{1ex}

where $F \odot G \defs (\dom(G) \ndres F) \oplus (\dom(F) \ndres G)$
\end{definition}
An external choice between two functions $F$ and $G$ essentially combines all the choices presented using $F \odot G$. The caveat is that if the domains of $F$ and $G$ overlap, then any events in common are excluded. Thus, $\odot$ restricts the domain of $F$ to maplets $e \mapsto P$ where $e \notin \dom(G)$, and vice-versa. This has the effect that $(a \then P) \extchoice (a \then Q) = \skey{stop}$, for example. In the special case that $\dom(F) \cap \dom(G) = \emptyset$, $P \odot Q = P \oplus Q$. We chose this behaviour to ensure that $\extchoice$ is commutative, though we could alternatively bias one side.

Internal steps on either side of $\extchoice$ are greedily consumed. Due to the equation order, $\tau$ events have the highest priority, following a maximal progress assumption~\cite{Hennessy1995TPL}. Return events also have priority over visible events. If two returns are present then they must agree on the value, otherwise they deadlock. External choice satisfies several properties: \isalink{https://github.com/isabelle-utp/interaction-trees/blob/df092d827c91393ea5b29a0cece4567380a8c931/ITree_CSP.thy\#L231}
  $$P \extchoice Q = Q \extchoice P \quad \skey{stop} \extchoice P = P \quad \skey{div} \extchoice P = \skey{div} \quad P \extchoice (\tau^n~Q) = (\tau^n~P) \extchoice Q = \tau^n (P \extchoice Q)$$

  \vspace{-5ex}
  
  $$(\Vis~F \extchoice \Vis~G) \mbind H = (\Vis~F \mbind H) \extchoice (\Vis~G \mbind H)$$
External choice is commutative and has \skey{stop} as a unit. It has \skey{div} as an annihilator, because the $\tau$ events means that no other activity is chosen. A finite number of $\tau$ events on either the left or right can be extracted to the front. Finally, bind distributes from the left across a visible event choice. We prove these properties using coinduction (\cref{thm:coind}).

Using the operators defined so far, we can implement a simple buffer process: \isalink{https://github.com/isabelle-utp/interaction-trees/blob/7695143479e6f604209545c500db1c6ee6d25faa/examples/ITree_CSP_Examples.thy\#L23}
  
\begin{alltt}
\isakwmaj{chantype} Chan = Input::integer  Output::integer  State::"integer list"

\isakwmaj{definition} buffer :: "integer list \(\Rightarrow\) (Chan, integer list) itree" \isakwmin{where}
"buffer = loop (\(\lambda\) s. 
                 do \{ i \(\leftarrow\) inp Input \{0..\}; Ret (s @ [i]) \}
               \(\extchoice\) do \{ guard(length s > 0); outp Output (hd s); Ret (tl s) \}
               \(\extchoice\) do \{ outp State s; Ret s \})"
\end{alltt}
We first create a channel type \texttt{Chan}, which has channels (prisms) for inputs and outputs, and to view the current buffer state. We define the buffer process as a simple loop with a choice with three branches inside. The variable \texttt{s::integer list} denotes the state. The first branch allows a value to be received over \texttt{Input}, and then returns \texttt{s} with the new value added, and then iterates. The second branch is only active when the buffer is not empty. It outputs the head on \texttt{Output}, and then returns the tail. The final branch simply outputs the current state. In \S\ref{sec:simulation} we will see how such an example can be simulated.

Next, we tackle parallel composition. The objective is to define the usual CSP operator $P \parallel[E] Q$, which requires that $P$ and $Q$ synchronise on the events in $E$ and can otherwise evolve independently. We first define an auxiliary operator for merging choice functions.
\begin{align*}
   merge_E(F, G) &= (\lambda e \in \dom(F) \setminus (\dom(G) \cup E) @ \skey{Left}(F(e))) \\
                 &\,\oplus (\lambda e \in \dom(G) \setminus (\dom(F) \cup E) @ \skey{Right}(G(e))) \\
                 &\,\oplus (\lambda e \in \dom(F) \cap \dom(G) \cap E @ \skey{Both}(F(e), G(e))
\end{align*}
Operator $merge_E(F, G)$ merges two event functions. Each event is tagged depending on whether it occurs on the $\skey{Left}$, $\skey{Right}$, or $\skey{Both}$ sides of a parallel composition. An event in $\dom(F)$ can occur independently when it is not in $E$, and also not in $\dom(G)$. The latter proviso is required, like for $\extchoice$, to prevent nondeterminism by disallowing the same event from occurring independently on both sides. An event in $\dom(G)$ can occur independently through the symmetric case with for $\dom(F)$. An event can synchronise provided it is in the domain of both choice functions and the set $E$. We use this operator to define generalised parallel composition. For the sake of presentation, we present partial functions as sets.

\begin{definition} $P \parallel_E Q$ is defined corecursively by the following equations: \isalink{https://github.com/isabelle-utp/interaction-trees/blob/7695143479e6f604209545c500db1c6ee6d25faa/ITree_CSP.thy\#L321}
\begin{align*}
    (\Vis~F) \parallel_E (\Vis~G) &= 
        \Vis\left(\begin{array}{l}
            \{e \mapsto (P' \parallel_E (\Vis~G)) | (e \mapsto \skey{Left}(P')) \in merge_A(F, G)\} \\
            \oplus~ \{e \mapsto ((\Vis~F) \parallel_E Q') | (e \mapsto \skey{Right}(Q')) \in merge_E(F, G)\} \\
            \oplus~ \{e \mapsto (P' \parallel_E Q') | (e \mapsto \skey{Both}(P', Q')) \in merge_E(F, G)\}
        \end{array}\right) \\
    (\Sil~P') \parallel_E Q &= \Sil~(P' \parallel_E Q) \qquad P \parallel_E (\Sil~Q') = \Sil~(P \parallel_E Q') \\
    (\Ret~x) \parallel_E (\Ret~y) &= \Ret~(x, y) \\
    (\Ret~x) \parallel_E (\Vis~G) &= \Vis~\{e \mapsto \Ret~x \parallel_E Q' | (e \mapsto Q') \in G\} \\
    (\Vis~F) \parallel_E (\Ret~y) &= \Vis~\{e \mapsto P' \parallel_E \Ret~y | (e \mapsto P') \in F\}
\end{align*}

\end{definition}
The most complex case is for $\Vis$, which constructs a new choice function by merging $F$ and $G$. The three cases are again represented by three partial functions. The first two allow the left and right to evolve independently to $P'$ and $Q'$, respectively, using one their enabled events, leaving their opposing side, $\Vis~G$ and $\Vis~F$ respectively, unchanged. The third case allows them both to evolve simultaneously on a synchronised event.

The $\Sil$ cases allow $\tau$ events to happen independently and with priority. If both sides can return a value, $x$ and $y$, respectively then the parallel composition returns a pair, which can later be merged if desired. The final two cases show what happens when only one side has a return value, and the other side has visible events. In this case, the $\Ret$ value is retained and push through the parallel composition, until the other side also terminates.

We use $\parallel_E$ to define two special cases for CSP: $P \parallel[E] Q \defs (P \parallel_E Q) \mbind (\lambda (x, y) @ \Ret~())$ and $P \interleave Q \defs P \parallel[\emptyset] Q$.  As usual in CSP, these operators do not return any values and so $P, Q :: (E, ())\skey{itree}$.  The $P \parallel[E] Q$ operator is similar to $\parallel_E$, except that if both sides terminate any resultant values are discarded and a null value is returned. This is achieved by binding to a simple merge function. $P$ and $Q$ do not return values, and so this has no effect on the behaviour, just the typing. The interleaving operator $P \interleave Q$, where there is no synchronisation, is simply defined as $P \parallel[\emptyset] Q$. We prove several algebraic laws:
\isalink{https://github.com/isabelle-utp/interaction-trees/blob/7695143479e6f604209545c500db1c6ee6d25faa/ITree_CSP.thy\#L476}
  $$(P \parallel_E Q) = (Q \parallel_E P) \mbind (\lambda (x, y) @ \Ret~(y, x)) \quad \skey{div} \parallel_E P = \skey{div}$$
  $$P \parallel[E] Q = Q \parallel[E] P \quad P \interleave Q = Q \interleave P \quad \skey{skip} \interleave P = P$$
Parallel composition is commutative, except that we must swap the outputs, and so $\parallel[E]$ and
$\interleave$ are as well. Parallel has $\skey{div}$ as an annihilator for similar reasons to
$\extchoice$. For $\interleave$, $\skey{skip}$ is a unit since there is no possibility of communication and no values
are returned.

The final operator we consider is hiding, $P \hide A$, which turns the events in $A$ into $\tau$s:

\begin{definition}[Hiding] $P \hide A$ is defined corecursively by the following equations: \isalink{https://github.com/isabelle-utp/interaction-trees/blob/7695143479e6f604209545c500db1c6ee6d25faa/ITree_CSP.thy\#L571}
\begin{align*}
    \Vis(F) \hide A &= 
        \begin{cases}
            \Sil~ (F(e) \hide A) & \text{if } A \cap \dom(F) = \{e\} \\
            \Vis~ \{(e, P \hide A) | (e, P) \in F\} & \text{if } A \cap \dom(F) = \emptyset\\
            \skey{stop} & \text{otherwise}
        \end{cases} \\
    \Sil(P) \hide A & = \Sil(P \hide A) \qquad \Ret~x \hide A = \Ret~x
\end{align*}

\end{definition}
We consider a restricted version of hiding where only one event can be hidden at a time, to avoid nondeterminism. When hiding the events of $A$ in the choice function $F$ there are three cases: (1) there is precisely one event $e \in A$ enabled, in which case it is hidden; (2) no enabled event is in $A$, in which case the event remains visible; (3) more than one $e \in A$ is enabled, and so we deadlock. We again impose maximal progress here, so that an enabled event to be hidden is prioritised over other visible events: $(a \then P \mathop{\vbar} b \then Q) \hide \{a\} = \tau P$, for example. In spite of the significant restrictions on hiding, it supports the common pattern where one output event is matched with an input event. Moreover, a priority can be placed on the order in which events are hidden, rather than deadlocking, by sequentially hiding events. Hiding can introduce divergence, as the following theorem shows: $(\skey{iter}~(\skey{sync}~e)) \hide {e} = \skey{div}$.

\subsection{Circus}

Whilst CSP processes can be parametrised to allow modelling state, there is no support for explicit state operators like assignment. The $do$ notation somewhat allows variables, but these are immutable and are not preserved across iterations. \Circus~\cite{Woodcock2001-Circus,Oliveira&09} is an extension of CSP that allows state variables. Given a state variable \texttt{buf::integer list}, the buffer example can be expressed in \Circus as follows:
\isalink{https://github.com/isabelle-utp/interaction-trees/blob/4bdea2d0a52341e7a19abc3950a3bcdd4b65e7fd/examples/ITree_Circus_Examples.thy\#L9}
\begin{align*}
buf := [] &\relsemi loop ((Input?(i) \then buf := buf \append [i]) \\
&     \qquad \extchoice ((length(buf) > 0) \guard Output!(hd~buf) \then buf := tl~buf) \\
&     \qquad \extchoice State!(buf) \then \skey{Skip})
\end{align*}
We update the state with assignments, which are threaded through sequential composition.

In our work~\cite{Foster17c,Foster2020-IsabelleUTP,Foster2021-JLAMP}, each state variable is modelled as a lens~\cite{Foster09}, $x :: \view \lto \src$. This is a pair of functions $\lget :: \view \to \src$ and $\lput :: \src \to \view \to \src$, which query and update the variables present in state $\src$, and satisfy intuitive algebraic laws~\cite{Foster2020-IsabelleUTP}. They allow an abstract representation of state spaces, where no explicit model is required to support the laws of programming~\cite{Hoare87}. Lenses can be designated as independent, $x \lindep y$, meaning they refer to different regions of $\src$. An expression on state variables is simply a function $e :: \src \to \view$, where $\view$ is the return type. We can check whether an expression $e$ uses a lens $x$ using unrestriction, written $x \unrest e$. If $x \unrest e$, then $e$ does not use $x$ in its valuation, for example $x \unrest\,(y + 1)$, when $x \lindep y$. Updates to variables can be expressed using the notation $[x_1 \leadsto e_1, x_2 \leadsto e_2, \cdots]$, with $x_i :: \view_i \lto \src$ and $e_i :: \src \to \view_i$, which represents a function $\src \to \src$.

We can characterise \Circus through a Kleisli lifting of CSP processes that return values, so that \Circus actions are simply homogeneous KTrees. We define the core operators below:
\begin{definition}[Circus Operators] \label{def:circus-ops} \isalink{https://github.com/isabelle-utp/interaction-trees/blob/4bdea2d0a52341e7a19abc3950a3bcdd4b65e7fd/ITree_Circus.thy\#L7}
\begin{align*}
  \assigns{\sigma} &\defs (\lambda s @ \Ret(\sigma(s))) \\
  x := e &\defs \assigns{[x \leadsto e]} \\
  c?x{:}A \then F(x) &\defs (\lambda s @ \skey{inp}~c~A \mbind (\lambda x @ F(x)~s)) \\
  c!e \then P &\defs (\lambda s @ \skey{outp}~c~(e~s) \mbind (\lambda x @ P~s)) \\
  P \extchoice Q &\defs (\lambda s @ P(s) \extchoice Q(s)) \\
  P \sfpar{ns_1}{E}{ns_2} Q &\defs \left(\lambda s @ (P(s) \parallel_E Q(s)) \mbind (\lambda (s_1, s_2) @ \lovrd{\lovrd{s}{s_1}{ns_1}}{s_2}{ns_2})\right)
\end{align*}
\end{definition}
\noindent Operator $\assigns{\sigma}$ lifts a function $\sigma : \src \to \src$ to a KTree. It is principally used to represent assignments, which can be constructed using our maplet notation, such that a single assignment $x := e$ is $\assigns{[x \leadsto e]}$. Most of the remaining operators are defined by lifting of their CSP equivalents. An output $c!e \then P$ carries an expression $e$, rather than a value, which can depend on the state variables. The main complexity is the \Circus parallel operator, $P \sfpar{ns_1}{E}{ns_2} Q$, which allows $P$ and $Q$ to act on disjoint portions of the state, characterised by the name sets $ns_1$ and $ns_2$. We represent $ns_1$ and $ns_2$ as independent lenses, $ns_1 \lindep ns_2$, though they can be thought of as sets of variables with $ns_1 \cap ns_2 = \emptyset$. The definition of the operator first lifts $\parallel_E$, and composes this with a merge function. The merge function constructs a state that is composed of the $ns_1$ region from the final state of $P$, the $ns_2$ region from $Q$, and the remainder coming from the initial state $s$. This is achieved using the lens override operator $\lovrd{s_1}{s_2}{X}$, which extracts the region described by $X$ from $s_2$ and overwrites the corresponding region in $s_1$, leaving the complement unchanged.

Our \Circus operators satisfy many standard laws~\cite{Oliveira&09,Foster2021-JLAMP}, beyond the CSP laws: \isalink{https://github.com/isabelle-utp/interaction-trees/blob/4bdea2d0a52341e7a19abc3950a3bcdd4b65e7fd/ITree_Circus.thy\#L44}
  \begin{align*}
    \assigns{\sigma} \fatsemi \assigns{\rho} &~=~ \assigns{\rho \circ \sigma} \\
    \assigns{\sigma} \fatsemi (P \extchoice Q) &~=~ (\assigns{\sigma} \fatsemi P) \extchoice (\assigns{\sigma} \fatsemi Q) \\
    x := e \fatsemi y := f &~=~ y := f \fatsemi x := e & \text{if } x \lindep y, x \unrest f, y \unrest e \\
    P \sfpar{ns_1}{E}{ns_2} Q &~=~ Q \sfpar{ns_2}{E}{ns_1} P & \text{if } ns_1 \lindep ns_2
  \end{align*}
Composition of state updates $\sigma$ and $\rho$ entails their composition. State updates distribute through external choice from the left. Two variable assignments commute provided their variables are independent and their respective expressions do not depend on the adjacent variable. \Circus parallel composition is commutative, provided we switch the name sets.

This concludes our discussion of CSP and \Circus. In the next section, we consider the failures-divergences semantics.

\section{Linking to Failures-Divergences Semantics}
\label{sec:sem-links}

Here, we show how ITrees are related to the standard failures-divergences semantics of CSP~\cite{Brookes1984}. The utility of this link is to allow ITrees to act as a target of refinement. Existing mechanisations of the CSP set-based and relational semantics~\cite{Taha2020CSP-Isabelle,Foster2021-JLAMP} can be used to capture nondeterministic specifications, and ITrees provide implementations. %

In the failures-divergences model, a process is characterised by two sets: $F :: (E^\tick~\skey{list} \, \times \, \textit{E}~\skey{set})~\skey{set}$ and $D :: \power (E~\skey{list})$, which are, respectively, the set of failures and divergences. A failure is a trace of events plus a set of events that can be refused at the end of the interaction. A divergence is a trace of events that leads to divergent behaviour. A distinguished event $\tick \in Event$ is used as the final element of a trace to indicate that this is a terminating observation. Here, we show how to extract $F$ and $D$ from any ITree, and the operators of \S\ref{sec:csp-circus}.

We begin by giving a big-step operational semantics to ITrees, using an inductive predicate.

\begin{definition}[Big-Step Operational Semantics] $ $ \isalink{https://github.com/isabelle-utp/interaction-trees/blob/4bdea2d0a52341e7a19abc3950a3bcdd4b65e7fd/Interaction_Trees.thy\#L310}

\centering
\begin{tabular}{ccc}
\AxiomC{--\vphantom{$P \xrightarrow{tr} P'$}}
\UnaryInfC{$P \xrightarrow{[]} P$}
\DisplayProof
&
\AxiomC{$P \xrightarrow{tr} P'$}
\UnaryInfC{$\tau P \xrightarrow{tr} P'$}
\DisplayProof
&
\AxiomC{$e \in E$}
\AxiomC{$F(e) \xrightarrow{tr} P'$}
\BinaryInfC{$\left(\vbar\, x \in E @ F(x)\right) \xrightarrow{e \# tr} P'$}
\DisplayProof
\end{tabular}
\end{definition}
The relation $P \xrightarrow{tr} Q$ means that $P$ can perform the trace of visible events contained in the list $tr : E~\skey{list}$ and evolves to the ITree $Q$. This relation skips over $\tau$ events. The first rule states that any ITree may perform an empty trace ($[]$) and remain at the same state. The second rule states that if $P$ can evolve to $P'$ by performing $tr$, then so can $\tau P$. The final rule states that if $e$ is an enabled visible event, and $P(e)$ can evolve to $P'$ by doing $tr$, then the event choice can evolve to $P'$ via $e \# tr$, which is $tr$ with $e$ inserted at the head.
With these laws, we can prove the usual operational laws for sequential composition as theorems:

\begin{theorem}[Sequential Operational Semantics] $ $ \isalink{https://github.com/isabelle-utp/interaction-trees/blob/4bdea2d0a52341e7a19abc3950a3bcdd4b65e7fd/Interaction_Trees.thy\#L420}

  \centering
  \begin{tabular}{ccc}
  $\begin{array}{c}
     - \\[.5ex] \hline
     \skey{skip} \rightarrow \ret{()}
   \end{array}$
  &
  $\begin{array}{c}
     P \xrightarrow{tr} P' \\ \hline
     (P \mbind Q) \xrightarrow{tr} (P' \mbind Q)
   \end{array}$
  &
  $\begin{array}{c}
     P \xrightarrow{tr_1} \ret{x} \quad Q(x) \xrightarrow{tr_2} Q' \\ \hline
     (P \mbind Q) \xrightarrow{tr_1 \mathop{\text{@}} tr_2} Q'
   \end{array}$
  \end{tabular}
\end{theorem}
The $\skey{skip}$ process immediately terminates, returning $()$. If the left-hand side $P$ of $\mbind$ can evolve to $P'$ performing the events in $tr$, then the overall bind evolves similarly. If $P$ can terminate after doing $tr_1$, returning $x$, and the continuation $Q(x)$ can evolve over $tr_2$ to $Q'$ then the overall $\mbind$ can also evolve over the concatenation of $tr_1$ and $tr_2$, $tr_1 \append tr_2$, to $Q'$.

Often in CSP, one likes to show that there are no divergent states, a property called divergence freedom. It is captured by the following inductive-coinductive definition:

\begin{definition}[Divergence Freedom] \isalink{https://github.com/isabelle-utp/interaction-trees/blob/4bdea2d0a52341e7a19abc3950a3bcdd4b65e7fd/ITree_Divergence.thy\#L30}
  $$\begin{array}{cccc}
  \begin{array}{c}
     - \\ \hline \ret{x} \SEarrow \mathcal{R}
  \end{array} &
  \begin{array}{c}
     P \SEarrow \mathcal{R} \\ \hline \tau P \SEarrow \mathcal{R}
  \end{array} &
  \begin{array}{c}
    \ran(F) \subseteq \mathcal{R} \\ \hline \Vis~F \SEarrow \mathcal{R}
  \end{array} &
    \skey{div-free} \defs \bigcup \, \{ \mathcal{R} | \mathcal{R} \subseteq \{ P | P \SEarrow \mathcal{R} \} \}
    \end{array}
  $$  
\end{definition}
Predicate $P \SEarrow \mathcal{R}$ is defined inductively. It requires that $P$ stabilises to a $\Ret$, or to a $\Vis$ whose coninuations are all contained in $\mathcal{R}$. Then, \skey{div-free} is the largest set consisting of all sets $\mathcal{R} = \{P | P \SEarrow \mathcal{R}\}$, and is coinductively defined. If we can find an $\mathcal{R}$ such that for every $P \in \mathcal{R}$, it follows that $P \SEarrow \mathcal{R}$, that is $\mathcal{R}$ is closed under stabilisation, then any $P \in \mathcal{R}$ is divergence free. Essentially, $\mathcal{R}$ needs to enumerate the symbolic post-stable states of an ITree; for example $\mathcal{R} = \{\skey{run}~E\}$ satisfies the provisos and so $\skey{run}~E$ is divergence free. We have proved that $P \in \skey{div-free} \iff (\nexists s @ P \xrightarrow{s} \skey{div})$, which gives the operational meaning.

With our transition relation, we can define Roscoe's step relation, which is used to link the operational and denotational semantics of CSP~\cite[Section~9.5]{Roscoe2010-UCS}:
\isalink{https://github.com/isabelle-utp/interaction-trees/blob/4bdea2d0a52341e7a19abc3950a3bcdd4b65e7fd/ITree_FDSem.thy\#L46}
$$(P \xRightarrow{s} P') \defs ((\exists t \in \Sigma\,\skey{list} @ s = t \append [\ret{x}] \land P \xrightarrow{t} \ret{x} \land P' = \skey{stop}) \lor (set(s) \subseteq \Sigma \land P \xrightarrow{s} P'))$$
Here, $set(s)$ extracts the set of elements from a list. The step relation is similar to $\xrightarrow{s}$, except that the event type is adjoined with a special termination event $\ret{}$. We define the enlarged set $\Sigma^\checkmark \defs \Sigma \cup \{ \ret{x} | x \in \src\}$, which adds a family of events parametrised by return values, as in the semantics of Occam~\cite{Roscoe1984-Occam}, which derives from CSP. A termination is signalled when the transition relation reaches a $\Ret~x$ in the ITree, in which case the trace is augmented with $\ret{x}$ and the successor state is set to $\skey{stop}$. We often use a condition of the form $set(s) \subseteq \Sigma$ to mean that no $\ret{x}$ event is in $s$. We can now define the sets of traces, failures, and divergences~\cite{Roscoe2010-UCS}:

\begin{definition}[Traces, Failures, and Divergences] \isalink{https://github.com/isabelle-utp/interaction-trees/blob/4bdea2d0a52341e7a19abc3950a3bcdd4b65e7fd/ITree_FDSem.thy\#L77}
  \begin{align*}
    \skey{traces}(P) &\defs \{s | set(s) \subseteq \Sigma^\checkmark \land (\exists P' @ P \xRightarrow{s} P')\} \\
    P \mathop{\skey{ref}} E &\defs ((\exists F @ P = \Vis~F \land E \cap \dom(F) = \emptyset) \lor (\exists x @ P = \Ret~x \land \ret{x} \notin E)) \\
    \skey{failures}(P) &\defs \left\{(s, X) | set(s) \subseteq \Sigma^\checkmark \land (\exists Q @ P \xRightarrow{s} Q \land Q \mathop{\skey{ref}} X)\right\} \\
    \skey{divergences}(P) &\defs \{s \append t | set(s) \subseteq \Sigma \land set(t) \subseteq \Sigma \land (\exists Q @ P \xRightarrow{s} Q \land \divergent{Q})\}
  \end{align*}
\end{definition}
The set $\skey{traces}(P)$ is the set of all possible event sequences that $P$ can perform. For $\skey{failures}(P)$, we need to determine the set of events that an ITree is refusing, $P \mathop{\skey{ref}} E$. If $P$ is a visible event, $\Vis~F$, then any set of events $E$ outside of $\dom(F)$ is refused. If $P$ is a return event, $\Ret~x$, then every event other than $\ret{x}$ is refused. With this, we can implement Roscoe's form for the failures. Finally, the divergences is simply a trace $s$ leading to a divergent state $\divergent{Q}$, followed by any trace $t$. We exemplify these definitions with two calculations of failures:
\begin{align*}
  \skey{failures}(\skey{inp}~c~A) &=
    \begin{array}{l}
      \{([], E) | \forall x \in A @ c.x \notin E \} \cup \{([c.x], E) | x \in A \land \ret{} \notin E\} \\
      \cup~ \{([c.x, \ret{()}], E) | x \in A\}
    \end{array} \\[1ex]
  \skey{failures}(P \mbind Q) &=
    \begin{array}{l}
      \{(s, X) | set(s) \subseteq \Sigma \land (s, X \cup \{\ret{x} | x \in \src\}) \in \skey{failures}(P)\} \\
      \cup~ \{(s \append t, X) | \exists v @ s \append [\ret{v}] \in \skey{traces}(P) \land (t, X) \in \skey{failures}(Q(v))\}
    \end{array}
\end{align*}
The failures of $\skey{inp}~c~A$ consists of (1) the empty trace, where no valid input on $c$ is refused; (2) the trace where an input event $c.x$ occurred, and $\ret{()}$ is not being refused; and (3) the trace where both $c.x$ and $\ret{()}$ occurred, and every event is refused. The failures of $P \mbind Q$ consist of (1) the failures of $P$ that do not reach a return, and (2) the terminating traces of $P$, ending in $\ret{v}$ appended with a failure of $Q(v)$, the continuation.

We conclude this section with some important properties:

\begin{theorem}[Semantic Model Properties] \isalink{https://github.com/isabelle-utp/interaction-trees/blob/4bdea2d0a52341e7a19abc3950a3bcdd4b65e7fd/ITree_FDSem.thy\#L363}
  \begin{align*}
    & (s, X) \in \skey{failures}(P) \land (Y \cap \{x | s \append [x] \in \skey{traces}(P)\} = \emptyset) \implies (s, X \cup Y) \in \skey{failures}(P) \\
    & s \in \skey{divergences}(P) \land set(t) \subseteq \Sigma \implies s \append t \in \skey{divergences}(P) \\
    & P \approx Q \implies (\skey{failures}(P) = \skey{failures}(Q) \land \skey{divergences}(P) = \skey{divergences}(Q)) \\
    & P \in \skey{div-free} \iff \skey{divergences}(P) = \emptyset \\
    & P \in \skey{div-free} \implies (\forall s~a @ s \append [a] \in \skey{traces}(P) \implies (s, \{a\}) \notin \skey{failures}(P))
  \end{align*}
\end{theorem}
The first two are standard healthiness conditions of the failures-divergences model~\cite{Roscoe2010-UCS}, called \textit{\textbf{F3}} and \textit{\textbf{D1}}, respectively. \textit{\textbf{F3}} states that if $(s, X)$ is a failure of $P$ then any event that cannot subsequently occur after $s$, according to the \skey{traces}, must also be refused. \textit{\textbf{D1}} states that the set of divergences is extension closed. We have also proved that two weakly bisimilar processes have the same set of divergences and failures.  %

\section{Simulation by Code Generation}
\label{sec:simulation}

The Isabelle code generator~\cite{Haftman2010-CodeGen,Haftmann2013-DataRefinement} can be used to extract code from (co)datatypes, functions, and other constructs, to functional languages like SML, Haskell, and Scala. Although ITrees can be infinite, this is not a problem for languages with lazy evaluation, and so we can step through the behaviour of an ITree. Code generation then allows us to support generation of verified simulators, and provides a potential route to correct implementations.

The main complexity is a computable representation of partial functions. Whilst $A \pfun B$ is partly computable, all that we can do is apply it to a value and see whether it yields an output or not. For simulations and implementations, however, we typically want to determine a menu of enabled events for the user to select from. Moreover, calculation of a semantics for CSP operators like $\extchoice$ and $\parallel$ requires us to compute with partial functions. For this, we need a way of calculating values for functions $\dom$, $\dres$, and $\oplus$, which is not possible for arbitrary partial functions. Instead, we need a concrete implementation and a data refinement~\cite{Haftmann2013-DataRefinement}.

We choose associative lists as an implementation, $A \pfun B \approx (A \times B)~\skey{list}$, which limits us to finite constructions. However, it has the benefit of being serialisable and so makes the simulator easier to implement. More sophisticated implementations are possible, as the core theory of ITrees is separated from the code generation setup. To allow us to represent partial functions by associative lists, we need to define a mapping function:
\begin{alltt}
\isakwmaj{fun} pfun_alist :: "('a \(\times\) 'b) list \(\Rightarrow\) ('a \(\pfun\) 'b)" \isakwmin{where}
"pfun_alist [] = \{\(\mapsto\)\}" | "pfun_alist ((k,v) # f) = pfun_alist f \(\oplus\) \{k \(\mapsto\) v\}"
\end{alltt}
This recursive function converts an associative list to a partial function, by adding each pair in the list as a maplet. We generally assume that associative lists preserve distinctness of keys, however for this function keys which occur earlier take priority. With this function we can then demonstrate how the different partial function operators can be computed. We prove the following congruence equations as theorems in Isabelle/HOL.
\isalink{https://github.com/isabelle-utp/Z_Toolkit/blob/b51b75fa419fb69d33d542238238e6f692732c37/Partial_Fun.thy\#L700}
    \begin{align*}
        (\textit{pfun\_alist}~f) \oplus (\textit{pfun\_alist}~g) &= \textit{pfun\_alist}~(g \mathop{\text{@}} f) \\
        A \dres (\textit{pfun\_alist}~f) &= \textit{pfun\_alist}~(\textit{AList.restrict}~A~m) \\
        (\lambda x \in (\textit{set}~xs) @ f(x)) &= \textit{pfun\_alist}~(map~(\lambda k @ (k, f~k))~xs)
    \end{align*}
Override ($\oplus$) is expressed by concatenating the associative lists in reverse order. Domain restriction ($\dres$) has an efficient implementation in Isabelle, \textit{AList.restrict}, which we use. For a partial $\lambda$-abstraction, we assume that the domain set is characterised by a list ($\textit{set}~xs$). Then, a $\lambda$ term can be computed by mapping the body function $f$ over $xs$.

\begin{figure}
  
  \centering
    \framebox{
      \includegraphics[width=11cm]{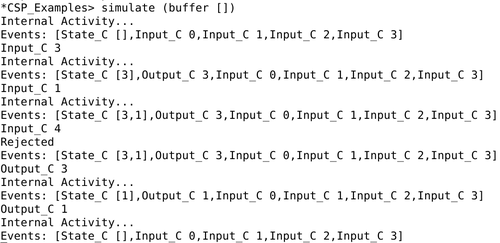}
      }

  \caption{Simulating the CSP buffer in the Glasgow Haskell Interpreter \hasklink{https://github.com/isabelle-utp/interaction-trees/blob/df092d827c91393ea5b29a0cece4567380a8c931/examples/Buffer_CSP.hs}}
  \label{fig:buffer}

  \vspace{-3ex}
\end{figure}

With these equations, we can set up the code generator. The idea is to designate certain representations of abstract types as code datatypes in the target language, of which each mapping function is a constructor. For sets, the following Haskell code datatype is produced:
\lstset{language=Haskell}
\begin{lstlisting}
data Set a = Set [a] | Coset [a] deriving (Read, Show);
\end{lstlisting}
A set is represented as a list of values using the constructor \texttt{Set}, which corresponds to the function $set$. It is often the case that we wish to capture a complement of another set, and so there is also the constructor \texttt{Coset} for a set whose elements are all those not in the given list. Functions on sets are then computed by code equations, which provide the implementation for each concrete representation. The membership function $member$ is implemented like this:

\begin{lstlisting}
member :: forall a. (Eq a) => a -> Set a -> Bool;
member x (Coset xs) = not (x `elem` xs); member x (Set xs) = xs `elem` x;
\end{lstlisting}
Each case for the function corresponds to a code equation. The function \lstinline{elem} is the Haskell prelude function that checks whether a value is in a list. This kind of representation ensures correctness of the generated code with respect to the Isabelle specifications. Similarly to sets, we can code generate the following representation for partial functions:
\hasklink{https://github.com/isabelle-utp/interaction-trees/blob/df092d827c91393ea5b29a0cece4567380a8c931/examples/Buffer_CSP.hs\#L36}

\begin{lstlisting}
data Pfun a b = Pfun_alist [(a, b)];

dom :: forall a b. Pfun a b -> Set a;
dom (Pfun_alist xs) = Set (map fst xs);
\end{lstlisting}
A partial function has a single constructor, although it is possible to augment this with additional representations. Each code equation likewise becomes a case for the corresponding recursive function, as illustrated by the domain function. Finally, we can code generate interaction trees, which are represented by a very compact datatype: \hasklink{https://github.com/isabelle-utp/interaction-trees/blob/df092d827c91393ea5b29a0cece4567380a8c931/examples/Buffer_CSP.hs\#L40}

\begin{lstlisting}
data Itree a b = Ret b | Sil (Itree a b) | Vis (Pfun a (Itree a b));
\end{lstlisting}
Each semantic definition, including corecursive functions, are also automatically mapped to Haskell functions. We illustrate the code generated for external choice below:
\hasklink{https://github.com/isabelle-utp/interaction-trees/blob/df092d827c91393ea5b29a0cece4567380a8c931/examples/Buffer_CSP.hs\#L173}

\begin{lstlisting}
extchoice :: (Eq a, Eq b) => Itree a b -> Itree a b -> Itree a b;
extchoice p q = (case (p, q) of {
    (Ret r, Ret y) -> (if r == y then Ret r else Vis zero_pfun);
    (Ret _, Sil qa) -> Sil (extchoice p qa); (Ret r, Vis _) -> Ret r;
    (Sil pa, _) -> Sil (extchoice pa q); (Vis _, Ret a) -> Ret a;
    (Vis _, Sil qa) -> Sil (extchoice p qa);
    (Vis f, Vis g) -> Vis (map_prod f g); });
\end{lstlisting}
The \lstinline{map_prod} function corresponds to $\odot$, and is defined in terms of the corresponding code generated functions for partial functions. The external choice operator ($\extchoice$) is simply defined as an infinitely recursive function with each of the corresponding cases in \cref{def:extchoice}.

For constructs like $\skey{inp}$ (\cref{def:basic-constructs}), there is more work to support code generation, since these can potentially produce an infinite number of events which cannot be captured by an associative list. Consider, for example, $\skey{inp}~c~\{0..\}$, for $c : \nat \pto E$, which can produce any event $c.i$ for $i \ge 0$. We can code generate this by limiting the value set to be finite, for example $\{0..3\}$. Then, the code generator maps this to a list $[0,1,2,3]$, which is computable. Thus, we can finally export code for concrete examples using the operator implementations.

We can now implement a simple simulator, the code for which is shown below: \hasklink{https://github.com/isabelle-utp/interaction-trees/blob/df092d827c91393ea5b29a0cece4567380a8c931/examples/Buffer_CSP.hs\#L222}

\begin{lstlisting}
sim_cnt :: (Eq e, Show e, Read e, Show s) => Int -> Itree e s -> IO ();
sim_cnt n (Ret x) = putStrLn ("Terminated: " ++ show x);
sim_cnt n (Sil p) = 
  do { if (n == 0) then putStrLn "Internal Activity..." else return ();
       if (n >= 20) 
       then do { putStr "Many steps (> 20); Continue?"; q <- getLine; 
                 if (q=="Y") then sim_cnt 0 p else putStrLn "Ended."; }
       else sim_cnt (n + 1) p };
sim_cnt n (Vis (Pfun_alist [])) = putStrLn "Deadlocked.";
sim_cnt n t@(Vis (Pfun_alist m)) = 
  do { putStrLn ("Events: " ++ show (map fst m)); e <- getLine;
       case (reads e) of
         []       -> do { putStrLn "No parse"; sim_cnt n t }
         [(v, _)] -> case (lookup v m) of
                       Nothing -> do { putStrLn "Rejected"; sim_cnt n t }
                       Just k -> sim_cnt 0 k };
simulate = sim_cnt 0;
\end{lstlisting}
The idea is to step through $\tau$s until we reach either a $\ret{x}$, in which case we terminate, or a $\Vis$, in which we case the user can choose an option. Since divergence is a possibility, we limit the number of $\tau$s that the will be skipped. After 20 $\tau$ steps, the user can choose to continue or abort the simulation. If an empty event choice is encountered, then the simulation terminates due to deadlock. Otherwise, it displays a menu of events, allows the user to choose one, and then recurses following the given continuation. The simulator currently depends on associative lists to represent choices, but other implementations are possible.

In order to apply the simulator, we need only augment the generated code for a particular ITree with the simulator code. \cref{fig:buffer} shows a simulation of the CSP buffer in \S\ref{sec:csp-circus}, with the possible inputs limited to $\{0.. 3\}$. We provide an empty list as a parameter for the initial state. The simulator tells us the events enabled, and allows us to pick one. If we try and pick a value not enabled, the simulator rejects this. Since lenses and expressions can also be code generated, we can also simulate the \Circus version of the buffer, with the same output.

As a more sophisticated example, we have implemented a distributed ring buffer, which is adopted from the original \Circus paper~\cite{Woodcock2001-Circus}. The idea is to represent a buffer as a ring of one-place cells, and a controller that manages the ring. It has the following form: \isalink{https://github.com/isabelle-utp/interaction-trees/blob/df092d827c91393ea5b29a0cece4567380a8c931/examples/RingBuffer.thy\#L96} $$(Controller \parallel[\{rd.c, wrt.c | c \in \nat\}] \left(\Interleave i\in\{0..maxbuff\} @ Cell(i))\right) \hide \{rd.c, wrt.c | c \in \nat\}$$ where $rd.c$ and $wrt.c$ are internal channels for the controller to communicate with the ring. Each cell is a single place buffer with a state variable $val$, and has the form $$Cell(i) \defs wrt?c \then val := v \relsemi loop (wrt?c \then val := v \extchoice rd!val \then \skey{Skip})$$ The cells are arranged through indexed interleaving, and $maxbuff$ is the buffer size. The channels $Input$ and $Output$ are used for communicating with the overall buffer. Space will not permit further details. The simulator can efficiently simulate this example, for a small ring with 5 cells, with a similar output to Figure~\ref{fig:buffer}, which is a satisfying result. \hasklink{https://github.com/isabelle-utp/interaction-trees/blob/df092d827c91393ea5b29a0cece4567380a8c931/examples/RingBuffer.hs}

We were also able to simulate the ring buffer with 100 cells, which requires about 3 seconds to compute the next step. With 1000 cells, the simulator takes more than a minute to calculate the next transition. The highest number of cells we could reasonably simulate is around 250. However, we have made no attempt to optimise the code, and several data types could be replaced with efficient implementations to improve scalability. Thus, as an approach to simulation and potentially implementation, this is very promising.

\section{Related Work}
\label{sec:related}

Infinite trees are a ubiquitous model for concurrency~\cite{Glabbeek1997CCS-CSP}. In particular, ITrees can be seen as a restricted encoding of Milner's synchronisation trees~\cite{Milner1980, Winskel1984STrees, Milner1989}. In contrast to ITrees, synchronisation trees allow multiple events from each node, including both visible and $\tau$ events. They have seen several generalisations, most recently by Ferlez et al.~\cite{Ferlez2014-GSTrees}, who formalise Generalized Synchronisation Trees based on partial orders, define bisimulation relations~\cite{Ferlez2018-BisimGSTrees}, and apply them to hybrid systems. Our work is different, because ITrees use explicit coinduction and corecursion, but there are likely mutual insights to be gained.

ITrees naturally support deterministic interactions, which makes them ideal for implementations. Milner extensively discusses determinism in \cite[chapter~11]{Milner1989}, a property which is imposed by construction in our operators. Similarly, Hoare defines a deterministic choice operator $a \then P | b \then Q$ in \cite[page~29]{Hoare85}, which is similar to ours except that Hoare's operator imposes determinism syntactically, where we introduce deadlock.

ITrees~\cite{ITrees2019}, and their mechanisation in Coq, have been 
applied in various projects as a way of defining abstract yet executable semantics~\cite{KLL+19,ZHHZ20,MHA20,ZZF20,ZHK+21,LPZ21,SZ21}. They have been used to verify C programs~\cite{KLL+19} and a HTTP key-value server~\cite{LPZ21}. The Coq mechanisation uses features not available in Isabelle, such as type constructor variables. Our novel mechanisation avoids the need for such features by fixing a universe for events, $E$, using partial functions to represent visible event choices, and using prisms~\cite{Pickering2017-Optics} to abstractly characterise channels.

\section{Conclusions}
\label{sec:concl}

In this paper we showed how Interaction Trees~\cite{ITrees2019} can be used to develop verified simulations for state-rich process languages with the help of Isabelle codatatypes~\cite{Blanchette2014BNF} and the code generator~\cite{Haftman2010-CodeGen, Haftmann2013-DataRefinement}. Our early results indicate that the technique provides both tractable verification, with the help of Isabelle's proof automation~\cite{Blanchette2016Hammers} and efficient simulation. We applied our technique to the CSP and \Circus process languages, though it is applicable to a variety of other process algebraic languages.

So far, we have focused primarily on deterministic processes, since these are easier to implement. This is not, however, a limitation of the approach. There are at least three approaches that we will investigate to handling nondeterminism in the future: (1) use of a dedicated indexed nondeterminism event; (2) extension of ITrees to permit a computable set of events following a $\tau$; (3) a further Kleisli lifting of ITrees into sets. Moreover, we will formally link ITrees to our formalisation of reactive contracts~\cite{Foster17c,Foster2021-JLAMP}, which provide a refinement calculus for reactive systems, building on our link with failures-divergences. We will implement the remaining CSP operators, such as renaming and interruption. We will also further investigate the failures-divergence semantics of our ITree process operators, and determine whether failures-divergences equivalence entails weak bisimulation.

Our work has many practical applications in production of verified simulations. We intend to use it to mechanise a
semantics for the RoboChart~\cite{Miyazawa2019-RoboChart} and RoboSim~\cite{Cavalcanti2019RoboSim} languages, which are
formal UML-like languages for modelling robots with denotational semantics based in CSP. This will require us to
consider discrete time, which we believe can be supported using a dedicated time event in ITrees, similar to
tock-CSP~\cite{Roscoe2005-TPC}. This will build on our colleagues' work with
$\checkmark$-$\textit{tock}$~\cite{Baxter2021-TickTock}, a new semantics for tock-CSP. This will open up a pathway from
graphical models to verified implementations of autonomous robotic controllers. In concert with this, we will also
explore links to our other theories for hybrid systems~\cite{Foster2020-dL,Foster19b-HybridRelations}, to allow
verification of controllers in the presence of a continuously evolving environment.

\bibliography{references}

\end{document}